\documentclass{amsart}

\usepackage{citesort}
\usepackage{a4wide}
\usepackage{amssymb}
\usepackage[dvips,final]{graphicx}

\begin{document}

\bibliographystyle{siam}

\title[Metastability and dispersive shock waves in Fermi-Pasta-Ulam
system]{Metastability and dispersive shock waves\\in Fermi-Pasta-Ulam system}

\author[P. Lorenzoni]{Paolo Lorenzoni}
\author[S. Paleari]{Simone Paleari}
\address{Universit\`a degli Studi di Milano~Bicocca\\
Dipartimento di Matematica e Applicazioni\\
Via R.~Cozzi, 53, 20125 --- Milano, Italy}
\email{paolo.lorenzoni@unimib.it}
\email{simone.paleari@unimib.it}

%\subjclass{??x?? (??x??, ??x??)}
\keywords{Metastability, dispersive shock waves, Fermi--Pasta--Ulam model}

\date{\today}

\begin{abstract}
We show the relevance of the dispersive analogue of the shock waves in
the FPU dynamics.  In particular we give strict numerical evidences
that metastable states emerging from low frequency initial
excitations, are indeed constituted by dispersive shock waves
travelling through the chain.  Relevant characteristics of the
metastable states, such as their frequency extension and their time
scale of formation, are correctly obtained within this framework,
using the underlying continuum model, the KdV equation.
\end{abstract}

\maketitle

%%%%%%%%%%%%%%%%%%%%%%%%%%%%%%%%%%%%%%%%%%%%%%%%%%%%%%%%%%%%%%%%%%%%%%%%%%%

\section{Introduction}

The Fermi--Pasta--Ulam (FPU) model was introduced in~\cite{FPU55} in
order to study the rate of relaxation to the equipartition in a chain
of nonlinearly interacting particles, starting with initial conditions
far away from equilibrium.  Both the original and the subsequent
numerical experiments show that this transition, if present, takes
much longer times than expected, raising the so called FPU problem, or
FPU paradox.

Recently there has been a growing interest in the field of continuum
limits and modulation equations for lattice models (see
e.g.,~\cite{FilV99,SchW00,FriP1-4,DreHM05,DreH05}, just to cite a few
without any claim of completeness).  Actually, it is possible to say
that one of the starting points of this activity goes back to a paper
by Zabusky and Kruskal~\cite{ZabK65}, where the recurrence phenomena
observed in the FPU system are related to the soliton behaviour
exhibited by the KdV equation, which emerges as a continuum limit of
the FPU.  Surprisingly, these two viewpoints evolved for several years
somewhat independently, with some people working essentially on
integrable systems, and other people interested mainly in the nearly
integrable aspects of the FPU type models.

In the first community the work~\cite{ZabK65} was a milestone.  A
crucial remark in that paper is that the time evolution of a class of
initial data for the KdV equation are approximately described by three
phases: one dominated by the nonlinear term, one characterized by a
balance between the nonlinear term and the dispersive term where some
oscillations, approximately described by a train of solitons with
increasing amplitude, start to develop, and the last where the
oscillations prevail.  Since the soliton velocities are related to
their amplitude, at a certain time they start to interact.  By
observing the trajectories of the solitons in the space-time, Kruskal
and Zabusky realized that, periodically, they almost overlap and
nearly reproduce the initial state.  Clearly, such an overlapping is
not exact, due to the phase shift given by the nonlinear interaction
between solitons.  They conjectured that this behaviour gives an
explanation of the FPU phenomenology.

As discovered later by Gurevich and Pitaevskii~\cite{GurP73} in a
different context, the oscillations observed by Zabusky and Kruskal
are more properly described in terms of modulated one-phase solutions
of KdV equations, and the slow evolution of their wave parameters is
governed by the so called Whitham's equations~\cite{Whi65}.

The analytic treatment of the dispersive analogue of shock waves is
less developed in the non integrable case even if some steps in this
direction have been made (see for instance \cite{El05}); moreover,
recently it has been conjectured that such phenomenon, in its early
development, is governed by a universal equation \cite{Dub05}.

From the FPU side, due to the original motivations arising from
Statistical Mechanics, the core question become the possible
persistence of the FPU problem in the thermodynamic limit, i.e. when
the number of particles $N$ goes to infinity at a fixed specific
energy (energy per particle).  Izrailev and Chirikov~\cite{IC66b}
suggested the existence of an energy threshold below which the
phenomenon is present; but such a threshold should vanish for growing
$N$ due to the overlapping of resonances; the recurrence in the FPU
system is thus explained in terms of KAM theorem (see
also~\cite{Rink01}).  According to Bocchieri et al.~\cite{BSBL}
instead, the threshold should be in specific energy and independent of
$N$.

Along the lines of the second conjecture, Galgani, Giorgilli and
coworkers~\cite{BeGG,BeGP,GPP,PP} have recently put into evidence a
metastability scenario (see also the previous paper by Fucito et
al.~\cite{FucitoP82}): for different classes of initial data, among
which those with low frequency initial conditions, the dynamics
reaches in a relatively short time a metastable state.  Such situation
can be characterized in terms of energy spectrum by the presence of a
natural packet of modes sharing most of the energy: and the
corresponding shape in the spectrum does not evolve for times which
appear to grow exponentially with an inverse power of the specific
energy.  The possible theoretical framework for such experimental
results could be a suitably weakened form of Nekhoroshev theory.

As one of the first attempts to join again in an original way the two
different aspects of the FPU paradox and of its integrable aspects, we
recall the paper~\cite{PogRK95}; in that work, the approaching of the
FPU chain to the shock is observed, but unfortunately the range of
specific energy  considered is too high to have the metastability (see
the higher part of the Fig.\ref{f.ms1}, left panel).

Some recent papers (see e.g., \cite{PonB05}) recovered the spirit of
the work by Zabuskii and Kruskal: exploiting the typical time scales
of the KdV equation, it is possible to recover those of the FPU
system, in particular the time scale of creation of the metastable
states; the length of the natural packet is also obtained and its
spectrum is expected to be connected to that of solitons of the
underlying KdV.

\medskip
Given all the elements we have put into evidence from the literature,
it is natural to make a further step.  Our claim is that, concerning
the metastability scenario, \emph{among the several aspects of the KdV
equation, precisely the dispersive analogue of the shock wave is the
relevant one}.

Thus the aim of this paper is to show, at least numerically, that the
metastable state is completely characterized by the formation and
persistence of the modulated solutions of KdV equation and by their
self interactions.  We are indeed able to put into evidence exactly
such phenomena \emph{directly in the FPU dynamics}.

In fact, we observe the formation of dispersive shock waves in the FPU
chain: correspondingly the metastable state sets in.  Exploiting the
scaling properties of the KdV equation one obtains for the spatial
frequency (wave number) of the dispersive waves, which can be
naturally thought as an upper bound for the natural packet, the
dependence on the specific energy $\epsilon$, as observed in the
numerical experiments: $\omega\sim\epsilon^{1/4}$.

Similarly we can also explain the dependence of the formation time of
the metastable state on the number $N$ of particles and on the
specific energy $\epsilon$: $t\sim N\epsilon^{-1/2}+\epsilon^{-3/4}$
when energy is initially placed on a low frequency mode.

Although this is a kind of preliminary and qualitative investigation,
we think in this way to give a unified and synthetic picture of the
metastability scenario in the FPU system, providing the precise
dynamical mechanism leading to its creation.

\medskip
In what follows, after a short description of the FPU model, we recall
the main facts about the metastability scenario in the FPU (see
Sect.~\ref{s.ms}), and some elements of the Gurevich--Pitaevskii
phenomenon for KdV equation (see Sect.~\ref{s.wh}); we then show in
Sect.~\ref{s.wh_fpu} the deep relations between the two aspects.

\subsection{The model}
\label{s.model}
In the original paper~\cite{FPU55}, Fermi, Pasta and Ulam considered
the system given by the following Hamiltonian
\begin{equation*}
  H(x,y) = \sum_{j=1}^{N} \left[ \frac{1}{2}y_j^2 +
	V\bigl(x_{j+1}-x_j\bigr)\right] \ , \quad
  V(s) = \frac{1}{2}s^2 + \frac{\alpha}{3}s^3 + \frac{\beta}{4}s^4 \ , 
\end{equation*}
describing the one--dimensional chain of $N+2$ particles with fixed
ends, which are represented by $x_0=x_{N+1}=0$.  For the free
particles, $x_1,\ldots,x_N$ are the displacements with respect to the
equilibrium positions.  This is the FPU $\alpha,\beta$--model.

The normal modes are given by
$$
x_j = \sqrt{\frac{2}{N+1}} \sum_{k=1}^{N} q_k \sin\frac{jk\pi}{N+1}
        \ ,\quad
y_j = \sqrt{\frac{2}{N+1}} \sum_{k=1}^{N} p_k \sin\frac{jk\pi}{N+1}\ ,
$$
$(q_k,p_k)$ being the new coordinates and momenta.  The quadratic part
of the Hamiltonian in the normal coordinates is given the form
\begin{equation}
\label{ej}
	H_2 = \sum_{j=1}^{N}E_j\ ,\quad
	E_j = \frac{1}{2}\bigl(p_j^2+\omega_j^2 q_j^2\bigr)\ ,\quad
	\omega_j = 2\sin\frac{j\pi}{2(N+1)}\ ,	
\end{equation}
$E_j$ being the harmonic energies and $\omega_j$ the harmonic
frequencies.

Given the following limit, representing the equipartition,
\begin{equation*}
  \overline E_j = \frac{1}{T} \int_0^T E_j(t)dt \longrightarrow
	\epsilon:=\frac{E}{N} \qquad \forall j \ ,
\end{equation*}
the fundamental question of the FPU problem is how long it takes
to be reached, if ever.

{\sl Remark:} It is clearly possible to consider also the case with
periodic boundary conditions.  In the present paper we focus our
attention on the case of fixed ends.  We also point out that, although
all the pictures shown but the first one refer to the
$\alpha$--model, the main qualitative aspects are still valid in the
$\beta$ one.

{\sl Remark:} All the numerical result present here, with the
exception of those in Fig. \ref{ZKexp}, are obtained using a second
order symplectic integrator (Verlet's algorithm), with a typical time
step of $0.05$.

%%%%%%%%%%%%%%%%%%%%%%%%%%%%%%%%%%%%%%%%%%%%%%%%%%%%%%%%%%%%%%%%%%%%%%%%%%%

\section{Metastability in the FPU chain}
\label{s.ms}

Concerning the FPU dynamics, many of the recent numerical evidences
\cite{BeGG,BeGP,PP,GPP} support a metastability scenario according to
which, below a certain threshold in specific energy, the system is
characterized by a metastable state; it remains frozen in such a state
for a time which grows rapidly as the specific energy goes to zero.

\begin{figure}[t]
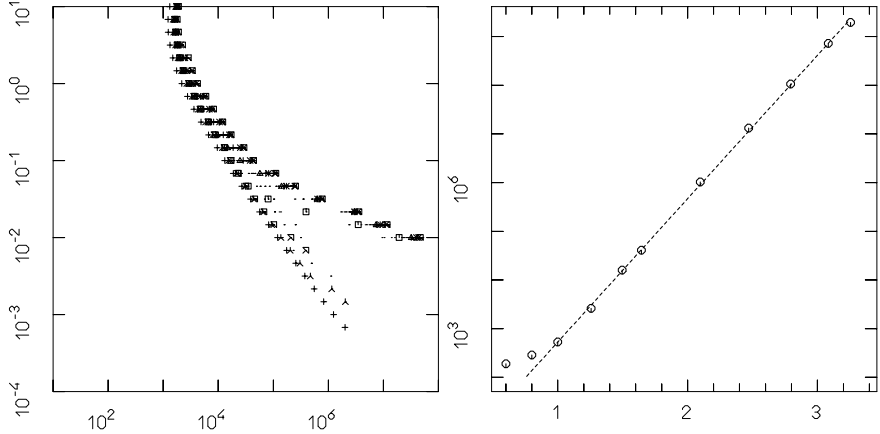

\begin{center}
\includegraphics[width=2.25truein]{Paleari_f_1.eps}   
\includegraphics[width=2.25truein]{Paleari_f_2.eps}   
\end{center}
\caption{Metastability.  Left panel: natural packet phenomenon.
$\beta$--model, with $\beta=1/10$, $N=255$, initial energy on mode
$1$; the threshold $\gamma$ is fixed to $0.1$.  Abscissa: critical
time $t_s$ for every packet$^1$; ordinate: specific energy
(from~\cite{PP}).  Right panel: exponentially long relaxation times.
$\beta$--model, with $\beta=1/10$.  Time needed by $n_{\rm eff}$ to
overcome the threshold $0.5$ vs. the specific energy power to $-1/4$,
for fixed $N=255$ and $\epsilon$ in the range $[0.0089,7.7]$, in
semi-log scale; every point is obtained averaging over 25 orbits with
different phases.  The straight line is the best fit using all the
points with $\epsilon\leq1$ (from~\cite{BeGP,PP}).}
\label{f.ms1}
\end{figure}

We will concentrate here on initial data with the whole energy in the
lowest frequency normal mode.  In order to better illustrate this
aspect of the dynamics, i.e. to show the time scales of these
metastable states and their dependence on the relevant parameters $E$
and $N$, we give in Fig.~\ref{f.ms1}, left panel, a typical picture.
Following \cite{BeGG}, one considers the energies $\mathcal
E_s=\sum_{j=1}^s E_j$ of packets of modes, and defines a corresponding
critical time $t_s$ as the time when the $s^{\rm th}$ packet has lost
a fixed fraction $\gamma$ of the energy it has to lose to reach
equipartition.  For every fixed specific energy, the different
critical times for all the possible packets are drawn in the picture.
From its qualitative aspect, which turns out to be practically
independent of the value of $\gamma$, one clearly recognizes two
different time scales, the first one giving the creation time of the
so called {\sl natural packet}, which then survive up to the second
time scale characterized by the reaching of the equipartition.  It has
been estimated, for the $\alpha,\beta$--model, a power law scaling of
the type $t\sim\epsilon^{-3/4}$ for the creation of the natural
packet, formed with normal modes with frequencies up to
$\omega\sim\epsilon^{1/4}$, with no dependence from $N$ in the range
$7-1\,023$ (see~\cite{BeGG}, Fig.~7).

As we said, the qualitative aspects of this figure (Fig.~\ref{f.ms1},
left panel) are present both in $\alpha$--model and $\beta$--model,
for different number of particles, and for initial data involving one
or more low frequency modes.

A second important point is the following: as observed from the very
beginning by Giorgilli and coworkers, the existence of these natural
packets appears to be an integrable phenomenon.  It is indeed possible
to produce the corresponding picture (figure not shown) for the Toda
lattice, which is integrable, and one obtains again the straight
branch of points corresponding to the creation of the packets, while
the second branch is completely missing, i.e. in the Toda chain these
packets seem to be stable structures.  The lower is the energy, the
longer the FPU takes to manifest its non integrability and to behave
differently from the Toda system.

A quantitative estimate of the second time scale related to the
equipartition is given in~\cite{BeGP} (but see also~\cite{PetL90}).
Using, as usual in this type of investigation, the spectral entropy
indicator $n_{\rm eff}:=\frac{e^S}N$, where $S:=-\sum_{j=1}^N e_j\ln
e_j$, and $e_j:=\frac{E_j}{\sum_{j=1}^N E_j}$, it is possible to
estimate the effective number of normal modes involved in the
dynamics.  Plotting the time necessary for $n_{\rm eff}$ to overcome a
fixed threshold close to its saturating value, as a function of the
specific energy, one obtains a reasonable estimate of the relaxation
times to equipartition, that is an estimate for the metastability
times.  The corresponding results are shown in Fig.~\ref{f.ms1}: the
numerical evidence supports the exponential scaling, with a possible
law of the type $T\sim\exp(\epsilon^{-1/4})$, for the $\beta$--model.

\begin{figure}[t]
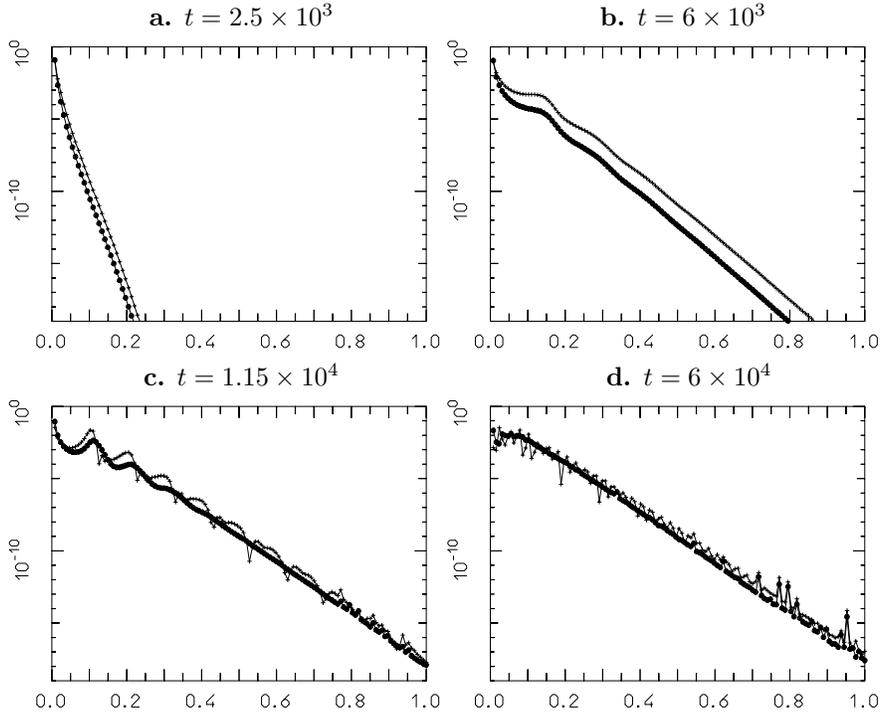

\hskip0.2in {\bf a.} $t=2.5\times10^3$ \hskip1.35in {\bf b.} $t=6\times10^3$
\smallskip
\begin{center}
\includegraphics[width=2.25truein]{Paleari_avg_a.eps}
\includegraphics[width=2.25truein]{Paleari_avg_b.eps}
\end{center}
\hskip0.2in {\bf c.} $t=1.15\times10^4$ \hskip1.35in {\bf d.} $t=6\times10^4$
\smallskip
\begin{center}
\includegraphics[width=2.25truein]{Paleari_avg_c.eps}
\includegraphics[width=2.25truein]{Paleari_avg_d.eps}
\end{center}
\caption{Time averaged distribution of energy among the harmonic modes
at different times.  $\alpha$--model, with $\alpha=1/4$, $N=127$,
specific energy $\epsilon=0.001$, initially on mode $1$.  Absciss\ae:
$k/N$, $k$ being the mode number.  Ordinates: the energy of the modes
averaged up to time $t$ reported in the label (thick points: average
from the beginning; thin points: average over a time window given by
the last 100 time units).
\label{f.avg}}
\end{figure}

The previous descriptions are concerned with a kind of more global
picture: concentrating on a single evolution at a fixed specific
energy, it is possible to have a better insight to the approach to
equipartition looking at the time evolution of the distribution of
energy among the modes.  We plot in Fig.~\ref{f.avg} some frames to
give an idea: the thick symbols represent the average over all the
times, while the thin ones are the averages over a short time window.
In the upper left panel, in the first stage of the evolution, one
observes the small but regular transfer of energy towards the higher
frequencies.  In the subsequent panel, as this process goes on, some
modulations on the shape of the spectrum appear.  At longer times the
spread to the higher part slows down, i.e. the slope of the straight
line converges to a certain value (see the lower left panel), and the
amplitude of the modulating bumps increases and their position slowly
moves leftward.  In the last frame we see quite well the spectral
structure of the metastable state: in the low frequency region the
slow motion of the bumps produces a plateau once one look at the
average over all times; and in the rest of the spectrum essentially
nothing happens a part from an extremely slow emergence of narrow
peaks.  Such a situation remain practically frozen for times growing
exponentially with a suitable power of the inverse of the specific
energy.

%%%%%%%%%%%%%%%%%%%%%%%%%%%%%%%%%%%%%%%%%%%%%%%%%%%%%%%%%%%%%%%%%%%%%%%%%%%
  
\section{The dispersive analogue of the shock waves}
\label{s.wh}
This section is a brief summary of some well-known facts about
dispersive shock waves. For our purposes it is sufficient to discuss
the case of the KdV equation
\begin{equation}
\label{KdV}
  u_t+uu_x+\mu^2 u_{xxx}=0
\end{equation}
which is known to give a good approximation the continuum limit of the
FPU model for long-wave initial conditions.

If the parameter $\mu$ is equal to zero it is well-known that, if the
initial data is decreasing somewhere, after a finite time $t_0$, the
gradient $u_x(x,t)$ of the solution becomes infinite in a point $x_0$
(point of gradient catastrophe).  For monotone initial data, the
solution is given in implicit form by the formula $x=tu+f(u)$, (where
$f(u)$ is the inverse function of the initial datum $u(x,0)$);
consequently, $u_x=1/(f'(u)+t)$ and the time of gradient catastrophe is
\begin{equation}
\label{ts}
  t_0=\max_{u\in\mathbb{R}}f'(u) \ .
\end{equation}

\begin{figure}[t]
%{\bf a.} $t=0$ \hskip1.2in {\bf b.} $t=0.27$ \hskip1.4in {\bf c.} $t=0.4$
\hfill{\bf a.} $t=0$\hfill\hfill {\bf b.} $t=0.27$\hfill\hfill {\bf c.} $t=0.4$\hfill
\smallskip
\begin{center}
\includegraphics[width=1.9truein]{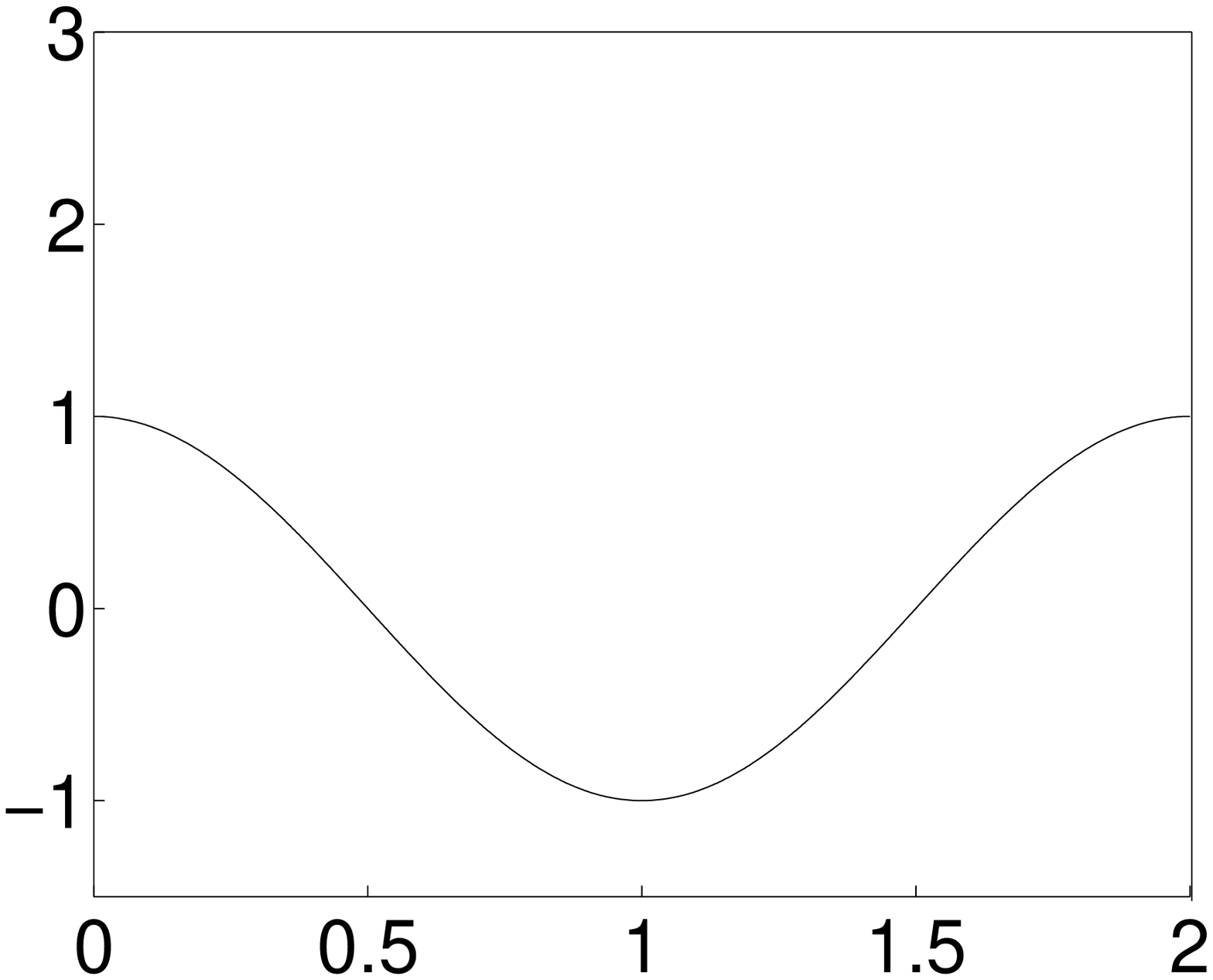}\hfill   
\includegraphics[width=1.9truein]{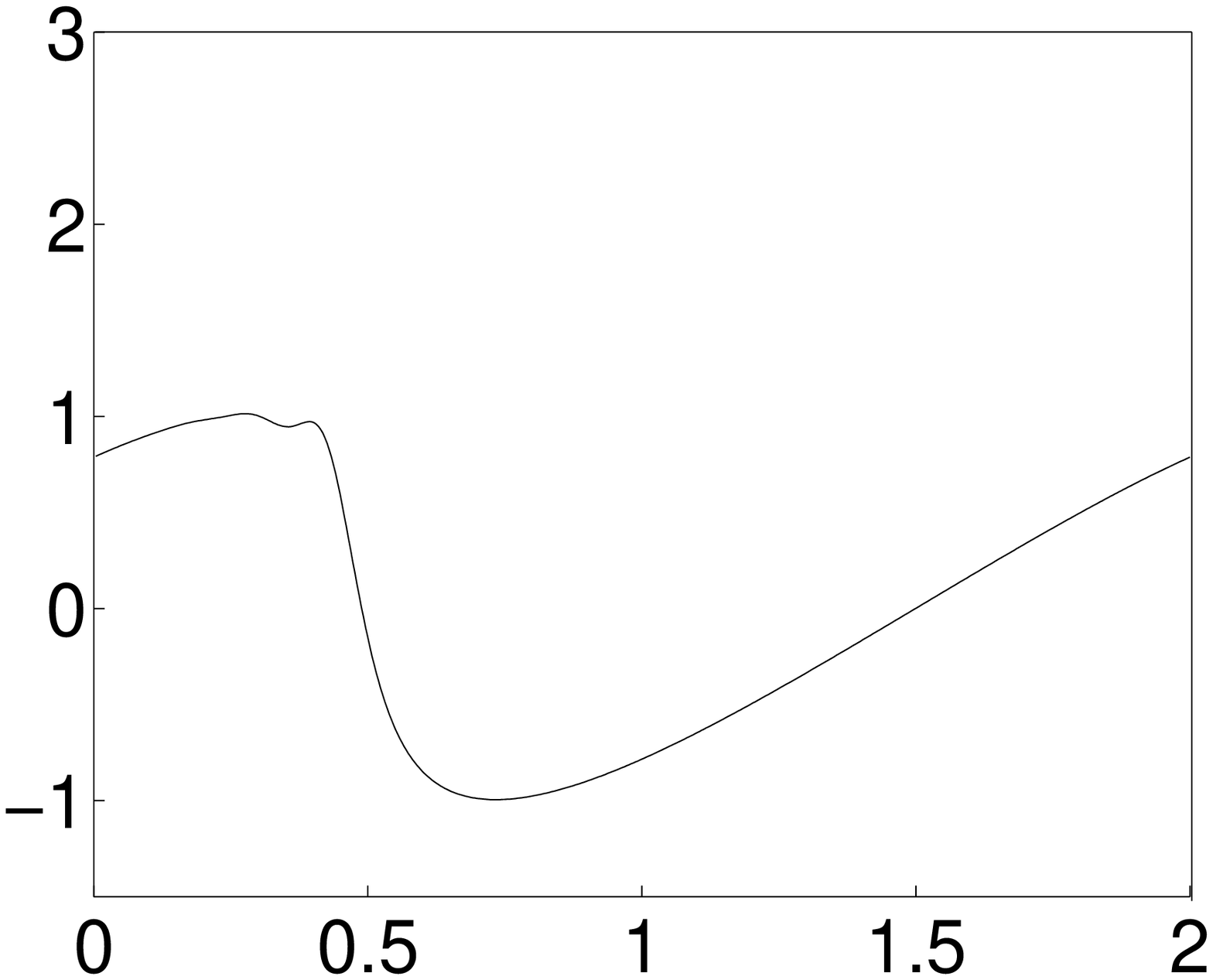}\hfill
\includegraphics[width=1.9truein]{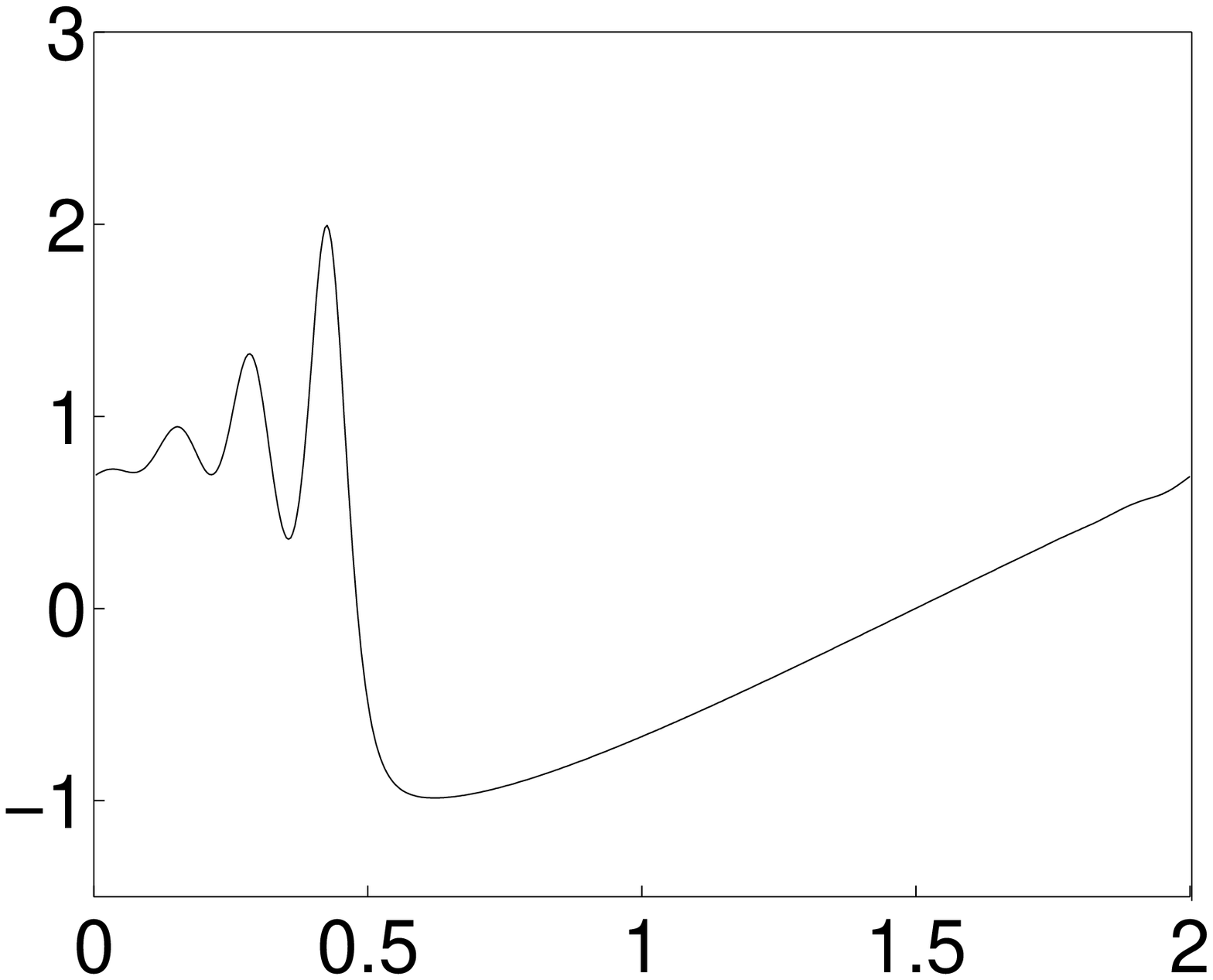}  
\end{center}
\caption{Formation of dispersive shock waves in the KdV equation
$u_t+uu_x+\mu^2u_{xxx}=0$, with $\mu=0.022$ (figures reproducing the
original Zabusky-Kruskal numerical experiments, obtained using a
Matlab code from~\cite{Sat05}).  The initial datum, shown in the left
panel, is $u(x,0)=\cos(\pi x)$; in the central panel the early stage
of formation of the dispersive shock is visible; in the right panel
one clearly observes the train of dispersive waves.}
\label{ZKexp}
\end{figure}

For $\mu\ne 0$, far from the point of gradient catastrophe, due to the
small influence of the dispersive term, the solution has a behaviour
similar to the previous case, while, in a neighbourhood of the singular
point, where the influence of the dispersion cannot be neglected
anymore, modulated oscillations of wave number of order $1/\mu$
develop (see Fig.\ref{ZKexp}).

According to the theory of Gurevich and Pitaevskii, these 
oscillations  can be approximately described in terms of one-phase 
solutions of the KdV equation
\begin{equation}
\label{ofs}
u(x,t)=u_2+a\, {\rm cn}^2\left(\sqrt{\frac{u_3-u_1}{3}}(x-ct);m\right)
\ ,
\end{equation}
that slowly vary in the time.

The wave parameters of the solution (\ref{ofs}) depend on the
functions $u_j$ which evolve according to an hyperbolic system of
quasi-linear PDEs: the Whitham's equations~\cite{Whi65}. Moreover, for
generic initial data, the region filled by the oscillations grows as
$t^{3/2}$~\cite{AviN87,Pot88,GurP73}.  Outside this region the
evolution of the solution is governed by the dispersionless KdV
equation.

Inside the oscillation region, two limiting situations arise when the
values of two of the three parameters $u_j$ coincide:

\begin{itemize}
\item $m\simeq 0$, $a\simeq 0$: in this case we have an harmonic wave
with small amplitude which corresponds to the small oscillations that
one can observe near the trailing edge of the oscillation zone;
\item $m=1$: in this case we have a soliton which one can observe near
the leading edge of the oscillation zone.
\end{itemize}

Since the Whitham's equations are hyperbolic, further points of
gradient catastrophe could appear in their solutions.  As a
consequence also multiphase solutions of KdV equation could enter into
the picture.
  
It turns out that, for $\mu\to 0$, the plane $(x,t)$ can be divided in
domains $D_g$ where the principal term of the asymptotics of the
solution is given by modulated $g$-phase solutions~\cite{LaxLD83}.
For monotonically decreasing initial data there exists rigorous
theorems which provide an upper bound to the number $g$~\cite{Gra04}.
Unfortunately for more general initial data, such as those we need,
the analytical description is less developed.

%%%%%%%%%%%%%%%%%%%%%%%%%%%%%%%%%%%%%%%%%%%%%%%%%%%%%%%%%%%%%%%%%%%%%%%%%%%

\section{Dispersive shock waves in the FPU}
\label{s.wh_fpu}

The connection between the FPU dynamics, when initial conditions
involve only low frequency modes, and KdV equation has been already
pointed out in the sixties by Zabusky and Kruskal~\cite{ZabK65}.  And
in many other subsequent papers it is heuristically deduced that the
KdV (respectively mKdV) may be viewed as continuum limit for FPU
$\alpha$--model (respectively $\beta$--model).

In this section we will first show in qualitative way how, in this
framework, the metastability is deeply related with the presence of
the dispersive shock waves; we will then give some more detail on the
continuum model involved.

\subsection{Metastability and dispersive shock waves}

As we described in Sec.~\ref{s.ms}, the metastability in the FPU model
is a property well visible in the energy spectrum (see
Fig.~\ref{f.avg}).  We show now that it can be also detected on the
dynamics of the particles.

\begin{figure}[t]
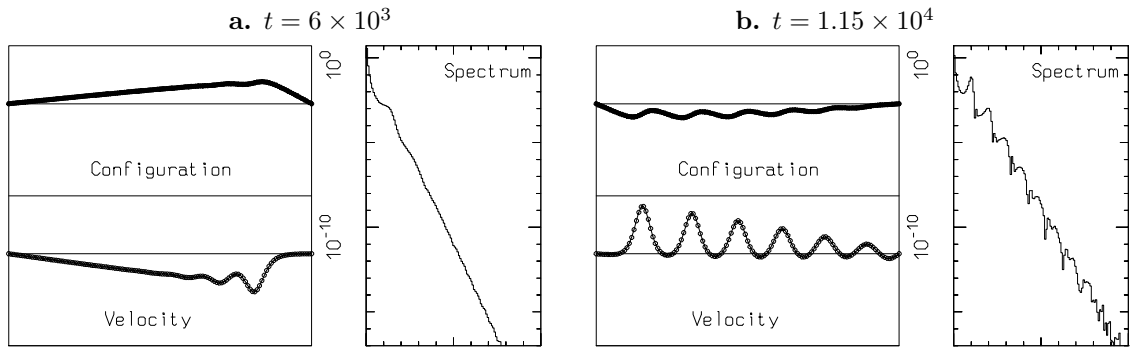

\hskip0.2in {\bf a.} $t=6\times10^3$ \hskip1.75in {\bf b.} $t=1.15\times10^4$
\smallskip
\begin{center}
\includegraphics[width=2.8truein]{Paleari_crd_a.eps}\hfill
\includegraphics[width=2.8truein]{Paleari_crd_b.eps}
\end{center}
\caption{Time evolution of the chain at different times, for the same
orbit as in fig.~\ref{f.avg}.  Absciss\ae: $k/N$, $k$ being the mode
number. Ordinates: as written on each subpanel.  The picture clearly
shows the correspondence between the appearance of the dispersive wave
and the emergence of bumps in the energy spectrum.
\label{f.crd}}
\end{figure}

In order to clarify this point we show in Fig.~\ref{f.crd} some frames
of the time evolution of positions and velocities of the chain.  We
choose the orbit whose spectrum is depicted in Fig.~\ref{f.avg}.

In the panel referring to time $t=6\times 10^3$, it is possible to see
that the emergence of the bump in the energy spectrum corresponds the
formation of the dispersive shock waves in the profile of the
velocity.  When the wave train has completely filled the chain, the
evolution of the energy spectrum stops: the system has reached the
metastable state (second panel).

It is worth to point out that the wave number of the modulated
waves is exactly of the same order of the bump; this fact suggests
that their slow motion, described in Sec.~\ref{s.ms}, could be
explained in terms of the evolution of the wave parameters (see
Eq.~\ref{ofs}) given by the Whitham's equations.

Thus the central point is that in the first stage of the dynamics, the
evolution may be divided in three regimes:

\smallskip
\noindent{\bf dispersionless regime:}

\begin{itemize}
\item[] the corresponding continuum model
is dominated by the non dispersive term, with the string approaching
the shock and the almost regular energy transfer towards higher
frequency modes, as shown in Fig.~\ref{f.avg}, panel a.
\end{itemize}

\smallskip
\noindent {\bf mixed regime:}
\begin{itemize}
\item[] in the corresponding PDE is visible the effect of the
dispersive term, which prevents the shock and creates the typical
spatial oscillations in the chain; correspondingly the energy spectrum
exhibits the formation of the bumps located at the correct frequency
and at its higher harmonics.  The part of the string non occupied by
the growing dispersive wave is still governed by the dispersionless
part and for this reason the flow of energy slows down and eventually
stops when the train involves the whole chain (see Fig.~\ref{f.avg},
panel b, and Fig.~\ref{f.crd}, panel a).
\end{itemize}

\smallskip
\noindent {\bf dispersive regime:}
\begin{itemize}
\item[] the dispersive shock wave fills the
lattice and travels through it: the metastable state is formed (see
Fig.~\ref{f.avg}, panel c, and Fig.~\ref{f.crd}, panel b).  In the
underlying integrable continuum limit one has the elastic interactions
of the solitons in the head of the train, together with the
recurrence phenomena already observed in~\cite{ZabK65}: this motivates
the long time persistence of the natural packet (see Fig.~\ref{f.avg},
panel d).
\end{itemize}

\smallskip

Such a correspondence between the evolution of the energy spectrum and
the appearance of dispersive waves and their growth up to the
occupation of the whole chain, has been observed in all our numerical
experiments: it appears to be independent of the number of particles
$N$, of the specific energy (in the range for which one has
metastability) and of the particular choice of the initial low
frequency mode excited.  We are working on a more systematic and
quantitative comparison.

In the remaining part of this section we will show more precisely that
the above mentioned dispersive and non dispersive regimes admit a clear
understanding in terms of the underlying continuum model.

\subsection{The continuum model}

As observed by Bambusi and Ponno~\cite{PonB05}, the FPU, with periodic
boundary conditions is well approximated, for low frequency initial
data, by a pair of evolutionary PDEs whose resonant normal form, in
the sense of canonical perturbation theory, is given by two
uncoupled KdV:
\begin{eqnarray}
\label{kdv1}
&&\xi_t=-\xi_x-\frac{1}{24N^2}\xi_{xxx}-\sqrt{\frac{\epsilon}{2}}\xi\xi_x
\ ,\\
\label{kdv2}
&&\eta_t=\eta_x+\frac{1}{24N^2}\eta_{xxx}+\sqrt{\frac{\epsilon}{2}}\eta\eta_x
\ ,
\end{eqnarray}
using the rescaling with respect to the original FPU variables
($x_{{\rm FPU}}= Nx_{{\rm KdV}}$, $t_{{\rm FPU}}= Nt_{{\rm KdV}}$);
$\xi$ and $\eta$ can be written in terms of two functions $q_x$ and
$p$ that, a part from a constant factor $\alpha$, respectively
interpolates the discrete spatial derivative $q_j-q_{j-1}$ and the
momenta $p_j$:
\begin{eqnarray*}
  &&\xi(x,t)\simeq \alpha(q_x+p)\\
  &&\eta(x,t)\simeq\alpha(q_x-p)
\end{eqnarray*}

\begin{figure}[t]
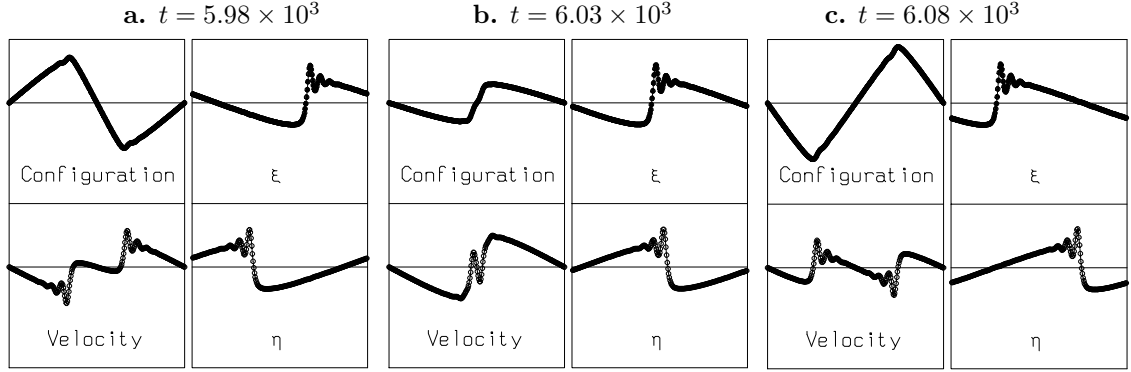

\hskip0.1in {\bf a.} $t=5.98\times10^3$ \hskip.75in {\bf b.} $t=6.03\times10^3$ \hskip.75in {\bf c.} $t=6.08\times10^3$
\smallskip
\begin{center}
\includegraphics[width=1.90truein]{Paleari_crd2_b.eps}\hfill
\includegraphics[width=1.90truein]{Paleari_crd2_c.eps}\hfill
\includegraphics[width=1.90truein]{Paleari_crd2_d.eps}
\end{center}
\caption{Apparent elastic interaction of dispersive wave trains,
explained in terms of the two decoupled KdV equations for $\xi$ and
$\eta$.  $\alpha$--model, with $\alpha=1/4$, $N=255$,
$\epsilon=0.001$, initial energy on mode $2$.  Absciss\ae: $k/N$, $k$
being the mode number.  In the velocity subpanel one observes two wave
trains travelling in opposite directions without interaction.  The
same two trains appear respectively in the $\xi$ and $\eta$ subpanels.
\label{f.crd2}}
\end{figure}

Taking into account that, in our case, we deal with fixed boundary
conditions, we must add to the equations (\ref{kdv1},\ref{kdv2}) the
following boundary conditions:
\begin{eqnarray*}
  &&\xi(0,t)=\eta(0,t) \\
  &&\xi(L,t)=\eta(L,t) \ .
\end{eqnarray*}
In other words the variables $\xi$ and $\eta$ exchange their role at
the boundary. Moreover, if $\eta(x,t)$ satisfies (\ref{kdv1}) then
$\eta(-x,t)$ satisfies (\ref{kdv2}) and vice-versa.

As a consequence, we can substitute the evolution of
$(\xi(x,t),\eta(x,t))$ with the evolution of a single function
$u(x,t)$ defined in terms of $\xi$ and $\eta$ in the following way
\begin{eqnarray*}
  &&u(x,t)=\xi(x,t) \hspace{1 cm}x\in [0,L] \\
  &&u(x,t)=\eta(2L-x,t) \hspace{1cm}x\in[L,2L] \ ,
\end{eqnarray*}
governed by the KdV equation
\begin{equation}
\label{uKdV}
  u_t=-u_x-\mu^2\frac{1}{24}u_{xxx}-\sqrt{\frac{\epsilon}{2}}uu_x \ ,
\end{equation}
and satisfying periodic boundary conditions.

\begin{figure}[t]
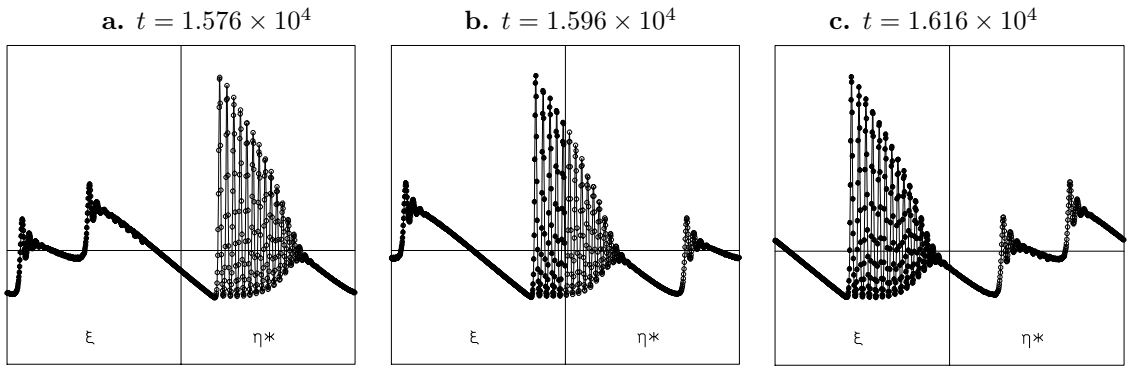

\hskip0.1in {\bf a.} $t=1.576\times10^4$ \hskip0.75in {\bf b.}
$t=1.596\times10^4$ \hskip.75in {\bf c.} $t=1.616\times10^4$
\smallskip
\begin{center}
\includegraphics[width=1.85truein]{Paleari_crd3_b.eps}\hfill
\includegraphics[width=1.85truein]{Paleari_crd3_c.eps}\hfill
\includegraphics[width=1.85truein]{Paleari_crd3_d.eps}
\end{center}
\caption{Evolution of the chain.  $\alpha$--model, with $\alpha=1/4$,
$N=511$, $\epsilon=0.001$, initial energy in equipartition on modes
$2$ and $3$.  Absciss\ae: the index particle $j$.  Ordinates: $\xi(j)$
in each left sub-panel (full circles), $\eta^*(j)=\eta(N-j)$ in each
right sub-panel (empty circles).  The dynamics develop three wave
trains; in these frames one observes the bigger one flowing in the
left direction and passing from the $\eta^*$ subpanel to the $\xi$
one, while the smaller ones go out the $\xi$ subpanel and reenter in
the other one: the two variables satisfy indeed a common KdV equation
on a circle.
\label{f.crd3}}
\end{figure}

These facts are clearly visible in Fig.~\ref{f.crd2} and
Fig.~\ref{f.crd3}.  In the first one, in order to show the role of the
variables $(\xi,\eta)$, we give three almost consecutive frames of the
evolution of a chain, with energy initially placed on the second
normal mode.  As expected, one has the generation of two trains of
dispersive waves (the number of trains generated is exactly $k$, when
the initially excited normal mode is the $k^{{\rm th}}$ one).  Looking
at the velocity sub-panel (the lower left one) of each frame, a sort
of elastic interaction appears between the trains: they approach each
other in frame (a), they interact in frame (b) and they come out
unmodified in frame (c).  But looking at the sub-panels of $\xi$ and
$\eta$ one realizes that such ``elastic interaction'' is the
superposition of two independent, in the normal form approximation,
dynamics.

Moreover, the relation between $(q,p)$ and $(\xi,\eta)$ explains why
the dispersive shock is well visible in the velocities, which are
directly a linear combination of the variables $\xi$ and $\eta$; the
configurations instead are approximately obtained by means of an
integration which has a smoothing effect.

Fig.~\ref{f.crd3} illustrates the role of fixed boundary conditions,
with the two KdV equations of the normal form for the periodic case,
degenerating into a single KdV.  In the three consecutive frames it is
perfectly clear that the big wave train supported by the $\eta$
variable flows into the equation for $\xi$, and the two smaller
trains, initially present in the $\xi$ subpanel, go out through the
left border and re-enter from the right side of the $\eta$ part.  The
true equation is the periodic KdV \ref{uKdV} for the variable $u$.

The recurrence appearing in the dispersive regime, and well visible in
our numerical experiments (figures not shown), may be seen as the
effect of the quasi-periodic motion on a torus of the underlying
integrable system.  On the other hand, following Kruskal and Zabusky,
it also has a natural explanation in terms of the elastic scattering
of solitons.  Clearly, a more refined description of this phenomenon
should also take into account the effect of the interactions of the
small oscillations near the trailing edge.  Indeed it could be
interesting to study the analytical aspects of the interaction of
dispersive wave trains, where the higher genus modulated solutions of
KdV equation could play a role.

\subsection{The relevant scalings}

In the previous section we showed that the continuum limit for the FPU
model with fixed ends, $N$ particles, specific energy $\epsilon$ and
an initial datum of low frequency, is given by the single KdV equation
\ref{uKdV}, which in the moving frame is
\begin{equation}
\label{mfKdV}
  u_t=-\frac{1}{24N^2}u_{xxx}-\sqrt{\frac{\epsilon}{2}}uu_x \ .
\end{equation}

Rescaling the space and time variables according to the following rule
\begin{eqnarray}
\label{r1}
  x \to N\epsilon^{1/4}x \ ,\\
\label{r2}
  t \to N\epsilon^{3/4}t \ ,
\end{eqnarray}
we get an equation independent of $N$ and $\epsilon$:
\begin{equation}
\label{risc}
  u_t=-\frac{1}{\sqrt{2}}uu_x-\frac{1}{24}u_{xxx} \ .
\end{equation}

We have seen that the metastable state is characterized by its size in
the modal energy spectrum, and by its time of creation and destruction.

Let us start with the first aspect.  Following the numerical evidence
we claim that the initial wave number of the dispersive wave provides
a good estimate, or at least an upper bound, for the width of the
packet of modes involved in the metastability.  Denoting by $\omega^*$
the relevant wave number for equation \ref{risc}, the corresponding
wave number $\omega_{{\rm KdV}}$ for equation \ref{mfKdV} is
\begin{equation*}
  \omega_{{\rm KdV}} = N\epsilon^{1/4} \omega^* \ ,
\end{equation*}
due to  the spatial rescaling \ref{r1}.

Recalling the spatial scaling in the continuum limit process ($x_{{\rm
FPU}}= Nx_{{\rm KdV}}$), one remains with the experimentally observed
law
\begin{equation*}
  \omega_{{\rm FPU}}\sim\epsilon^{1/4} \ ,
\end{equation*}
without any dependence on $N$.

Let us now consider the second point: the time scales.  As pointed out
in the previous discussion, the time of formation of metastable state
is characterized by two different regimes: one governed by
dispersionless KdV, where the effect of the nonlinearity prevails and
drive the dynamics towards a shock, and one governed by the full KdV
equation, where the dispersion prevents the shock and generates the
dispersive train.  Therefore we can approximately estimate the time of
formation of the metastable state as the sum of the two contribution
\begin{equation*}
  t_{{\rm FPU}} \simeq N(t_1+t_2) \ ;
\end{equation*} 
$t_1$ is the time of validity of the first regime, that we estimate
from above with the time $t_s$ of formation of the shock in the
dispersionless KdV, while $t_2$ is the time needed by the dispersive
wave to occupy the whole chain in equation \ref{mfKdV} ($t_2^*$ in
equation \ref{risc}), and the factor $N$ is again due to the rescaling
in the continuum limit process ($t_{{\rm FPU}}= Nt_{{\rm KdV}}$).

For the equation $u_t=\sqrt{\epsilon}uu_x$, one has
$t_s=\epsilon^{-1/2}F(u_0)$, where $F(u_0)$ is a function of the
initial data (see, e.g., the formula \ref{ts}).

Since $t_2=N^{-1}\epsilon^{-3/4}t_2^*$, due to the time rescaling
\ref{r2}, we obtain the law
\begin{equation*}
  t_{{\rm FPU}} \lesssim N\epsilon^{-1/2} + \epsilon^{-3/4} \ ;
\end{equation*}
we remark that such an estimate improves as $N$ grows because in that
case the first regime extends very close to the shock time due to the
delay in the dispersive effect.  Such a scaling fits with the
experiments~\cite[Fig. 8]{BeGG}.

\appendix

%%%%%%%%%%%%%%%%%%%%%%%%%%%%%%%%%%%%%%%%%%%%%%%%%%%%%%%%%%%%%%%%%%%%%%%%%%%

\section*{Acknowledgements}

We thank A.Giorgilli for useful discussions on the topic, and for
having introduced one of us in the study of the FPU model.

\bibliography{../biblio}

\end{document}